\documentclass[aps,pre,twocolumn,groupedaddress]{revtex4}
\usepackage{graphicx}
\usepackage{amsmath, amssymb}

\newcommand \be{\begin{eqnarray}}
\newcommand \ee{\end{eqnarray}}
\newcommand \ba{\begin{eqnarray}}
\newcommand \ea{\end{eqnarray}}
\newcommand \bs{\boldsymbol}
\newcommand \mc{\mathcal}
\newcommand \E{{\bs{\mc E}}}

\begin{document}
\title{Numerical convergence of the branching time of negative streamers}
\author{Carolynne Montijn$^1$, Ute Ebert$^{1,2}$, and Willem Hundsdorfer$^1$}
\affiliation{$^1$CWI, P.O.Box 94079, 1090 GB Amsterdam, The Netherlands,}
\affiliation{$^2$Dept.\ Physics, Eindhoven Univ.\ Techn., The Netherlands.}

\date{\today}

\begin{abstract}
In sufficiently large gaps and electric fields, discharge streamers 
do branch. In [Array\'as {\it et al.}, PRL {\bf 88}, 174502 (2002)], 
we observed streamer branching numerically within a deterministic 
particle density model and explained it as 
a Laplacian instability of a thin space charge layer. Our numerical
results were criticized in [Kulikovsky, PRL {\bf 89}, 229401 (2002)]. 
We here present an adaptive grid refinement method for streamer 
simulations, and we carry out the first conclusive investigation 
on the effect of the numerical grid on streamer branching in different 
fields. On stepwise finer grids the branching time converges, 
hence streamer branching is for the first time predicted quantitatively. 
\end{abstract}

\maketitle

{\bf Problem setting and review.}
Streamers are transient weakly ionized plasma channels 
that rapidly grow into a non- or weakly ionized medium
under influence of the self-enhanced electric field at their tip.
They are widely used in technology~\cite{vel2000,ebe2006} and ubiquitous 
in nature, where they play a role in creating the path of sparks, 
lightning~\cite{baz2000} and of blue jets above thunderclouds.
Streamers are also directly observed as so-called 
sprites~\cite{sen1995,ger2000,pas2002}, which are very large discharge structures in the 
higher parts of the atmosphere that are composed of ten thousands of streamers.
Despite their high velocity, streamer evolution is now directly observable
in experiments; a further review can be found in~\cite{ebe2006}.

Streamers commonly branch in experiments if gap and applied voltage 
are large enough.
Recently a debate has risen about the proper physical concept for
this branching. In 1939, Raether~\cite{rae1939} proposed a mechanism 
for streamer propagation and Loeb and Meek~\cite{loe1940} 
developed it into a branching concept that
nowadays is found in many textbooks. The concept is based 
on a uniformly charged streamer head; ahead of it stochastic 
processes create secondary avalanches, that subsequently develop 
into different branches. However, the distribution of rare electrons
due to photo-ionization or background ionization ahead of the streamer 
has never been shown to agree with the conceptual pictures, and the
concept has never been demonstrated to work.
Furthermore, simulations in the past two 
decades~\cite{dha1987,vit1994,kul1997-2,pan2001} have shown that
the fully developed streamer head is not homogeneously charged, 
but rather neutral and surrounded by a thin space charge layer which enhances
the field ahead of it and screens it in the interior; this field
enhancement allows the streamer to penetrate
regions with a rather low background field. 
Recent simulations also show that a streamer can branch within 
a fully deterministic model for charged particle densities, 
in a non-uniform background field~\cite{vit1993,hal2003,geo2005} as
well as in a uniform field~\cite{arr2002,roc2002,liu2004}, provided certain
requirements on the external parameters are met (e.g. a sufficiently
strong background electric field and a sufficiently long gap).

Some of the present authors have proposed~\cite{arr2002,roc2002} 
a physical explanation of these numerical observations that is directly 
related to the formation of the thin space charge layer: the layer
creates an almost equipotential streamer head that can undergo 
a Laplacian instability and branch in a manner similar 
to branching instabilities of fluid interfaces in viscous fingering. 
For a further discussion of the conceptual questions of streamer branching,
we refer to \cite{ebe2006}.
However, the numerical codes used in~\cite{vit1993,hal2003,geo2005,arr2002,
roc2002,liu2004} were not able to 
test the branching conditions on fine numerical grids. 
This lead some researchers
to question the physical nature of the instabilities~\cite{kul2002,pan2003,
hal2003, geo2005} despite the analytical arguments given 
in~\cite{arr2002,roc2002} and later in~\cite{meu2004,meu2005}.

To resolve the debate from the numerical side, we have developed
a code with comoving adaptive grids and we here present its results. 
The algorithm enables us to run the simulations on very fine grids;
therefore for the first time the effect of numerical grids 
on the branching process is investigated quantitatively. 
We here present its results: branching occurs both at very high fields 
like in Refs.~\cite{arr2002,roc2002} and also at fairly low 
background fields if the discharge has sufficient space to develop;
and the branching time saturates on sufficiently fine numerical grids.
This enables us to give the first quantitative predictions 
on streamer branching. 
\\

{\bf Model and multiscale structure of negative streamers.}
We investigate a minimal continuum model for streamers, 
which contains the essential physics for negative streamers 
in a non-attaching pure gas like N$_2$ or 
Ar~\cite{dha1987,vit1994,arr2002,roc2002}. The model is a two-fluid 
approximation for the charged particles, with a local field dependent 
impact ionization reaction coupled to the Poisson equation for 
electrostatic particle interactions. We investigate this 
model in a cylindrically symmetric geometry, reducing it to effectively
two dimensions. This constraint suppresses
one degree of freedom for the instability modes, and therefore the time of
branching in this cylindrical geometry is an upper bound for the branching 
time in a genuine three dimensional system~\cite{ebe2002-2,ebe2006}. 
In dimensionless units, the model reads
\ba
\displaystyle\partial_\tau\sigma 
  & = & \nabla\cdot(\sigma{\bs{\mathcal E}}+D\nabla\sigma)
        + \sigma|{\bs{\mathcal E}}|\;\alpha(|{\bs{\mathcal E}}|), 
\label{sigmaeq}\\[0mm]
\displaystyle\partial_\tau\rho
  & = & \sigma|{\bs{\mathcal E}}|\;\alpha(|{\bs{\mathcal E}}|), 
~~~\alpha(|{\bs{\mathcal E}}|) =e^{-1/|{\bs{\mathcal E}}|},
\label{rhoeq}\\[0mm]
\displaystyle - \nabla^2\phi& = &\rho-\sigma
  \, , \quad {\bs{\mathcal E}}=-\nabla\phi,
\label{phieq}
\ea
where $\sigma$ and $\rho$ are the electron and positive ion densities,
respectively. $\E$ and $\phi$ are respectively the electric field and 
potential, $D$ is the electron diffusion coefficient and $\tau$ is the
dimensionless time. The characteristic scales in this model depend
on the neutral gas density; therefore the simulation results 
can be applied to high altitude 
sprite discharges at low pressures as well as to high pressure laboratory 
experiments. We refer to~\cite{ebe1997,roc2002,ebe2006} for more details 
on the dimensional analysis.

A planar cathode is placed at $z=0$ and a planar anode at $z=L_z$.
The potential at the electrodes is fixed, $\phi(r,z=0,\tau)=0$,
$\phi(r,z=L_z,\tau)=\phi_0>0$, generating a background electric
field with strength $|\E_b|=\phi_0/L_z$ along the negative $z$-direction.
The streamer is initiated by an electrically neutral Gaussian ionization 
seed on the axis of symmetry at the cathode ($r=z=0$). There is no
background ionization far from the initial seed.

We impose homogeneous Neumann conditions for the electron density at 
all boundaries. This results in a net inflow of electrons from the cathode 
if the streamer is attached to it~\cite{arr2002,mon2005-2}. 
In practice, the computational volume is restricted in the radial
direction by a boundary $L_r$ sufficiently far away not to
disturb the solution near and in the streamer. Moreover, we choose the
inter-electrode distance $L_z$ so large that the streamer does not
feel the anode proximity for the results shown. 

The generic spatial structure of the streamer is already 
discussed above and can be seen in the figures: it
contains a wide range of spatial scales,
from the very extended non-ionized medium on which the Poisson equation 
has to be solved through the length of the conducting channel
and its width up to the inner structure of the thin space charge layer
around the streamer head.

Moreover, the region just ahead of the streamer where the field
is substantially enhanced and the electron density is low, 
is highly unstable, in the sense 
that a small ionized perturbation will grow much more rapidly 
than in the mere background field. This unstable region ahead
of the streamer tip is commonly referred to as {\em leading 
edge}~\cite{ebe1997,ebe2000}. It requires special 
care when considering numerical methods~\cite{ebe2000,mon2005-2}. 
Accurate simulations of streamers therefore pose a great computational
challenge.
\\

{\bf Numerical algorithm.}
In order to deal efficiently with the numerical challenges posed by this
model, it has been implemented in a numerical code using adaptive grid 
refinements. We recall the essential features of this algorithm 
and refer to \cite{mon2005-2} for further details. 
The spatial discretizations are based on finite volumes, using a flux 
limiting scheme to prevent spurious oscillations in the results near steep 
gradients.  The time stepping is performed with a two-stage explicit 
Runge-Kutta method.  

\begin{figure}
\begin{center}
\includegraphics[width=8cm]{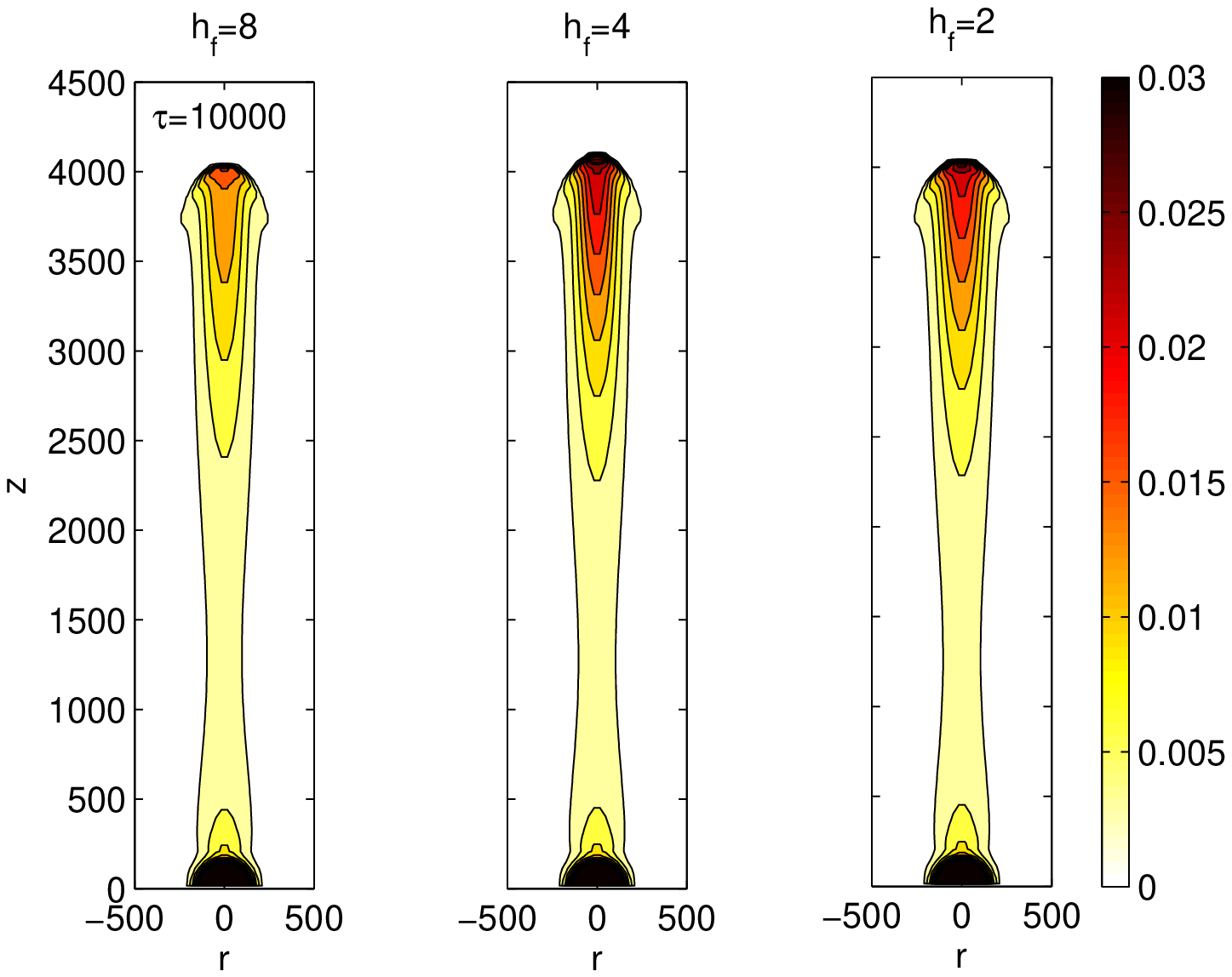}
\includegraphics[width=8cm]{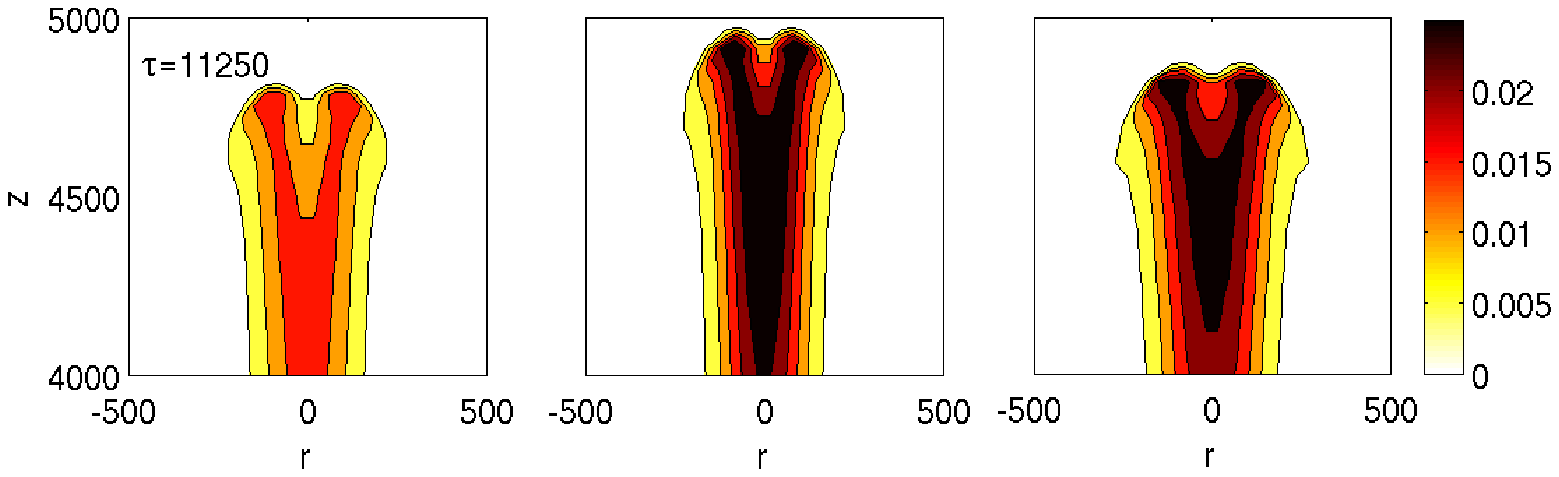}
\caption{Electron density distribution before and just after
streamer branching in a background field $|\E_b|=0.15$,
computed on different finest mesh sizes $h_f=$ 8, 4 and 2 as indicated 
over the plots. The upper snapshots at $\tau$=10000 are taken
before branching and the lower ones after branching, at time $\tau=11250$. 
The contours correspond to the same density levels. In all three cases 
the same restricted part of the total computational domain with 
$z\le L_z = 32768$ and $r\le L_r = L_z/2$ is shown.}
\label{br015}
\end{center}
\end{figure}

Using an explicit time-stepping method allows us to decouple the 
computational grids for the continuity 
equations~(\ref{sigmaeq})-(\ref{rhoeq}) on the one hand 
from those for the Poisson equation~(\ref{phieq}) on the other 
hand.  The particle densities are first updated on a series of nested, 
stepwise refined grids. Then the Poisson equation, using the computed 
densities as an input, is solved on another series of nested, stepwise 
refined grids. The electric field on the grids for the continuity equations 
is then calculated from the potential computed on the grids for the Poisson 
equation using sufficiently accurate interpolations~\cite{wac2005}.

\begin{figure}
\begin{center}
\includegraphics[width=8cm]{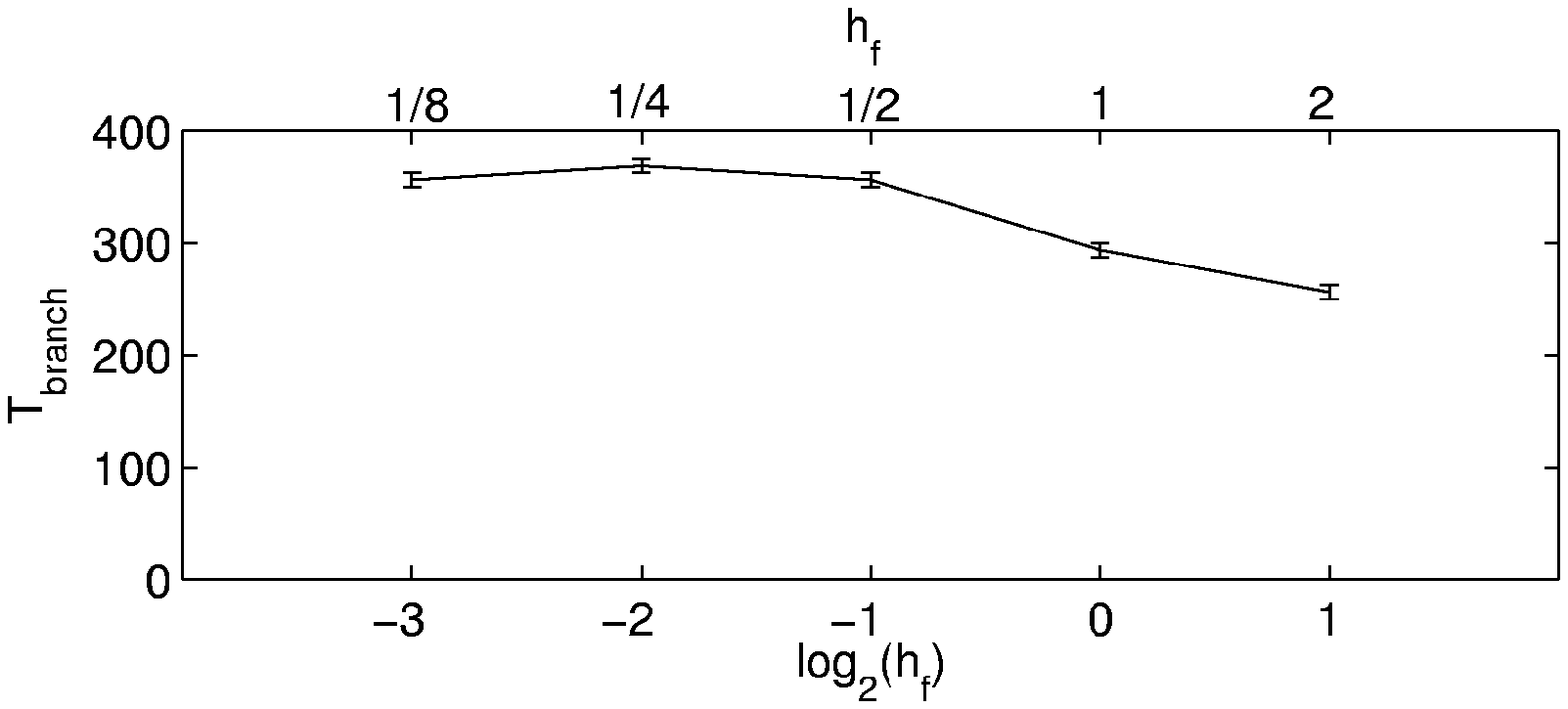}\\~\\
\includegraphics[width=8cm]{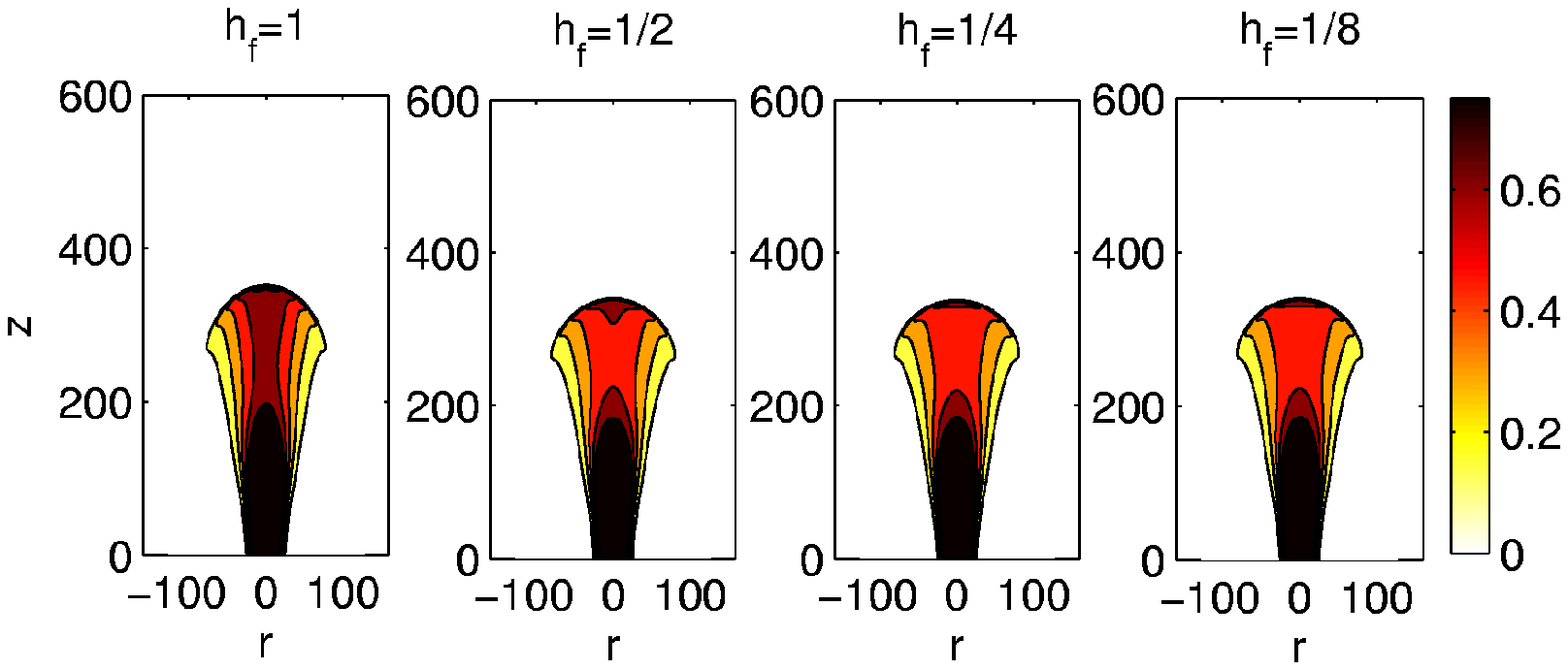}
\includegraphics[width=8cm]{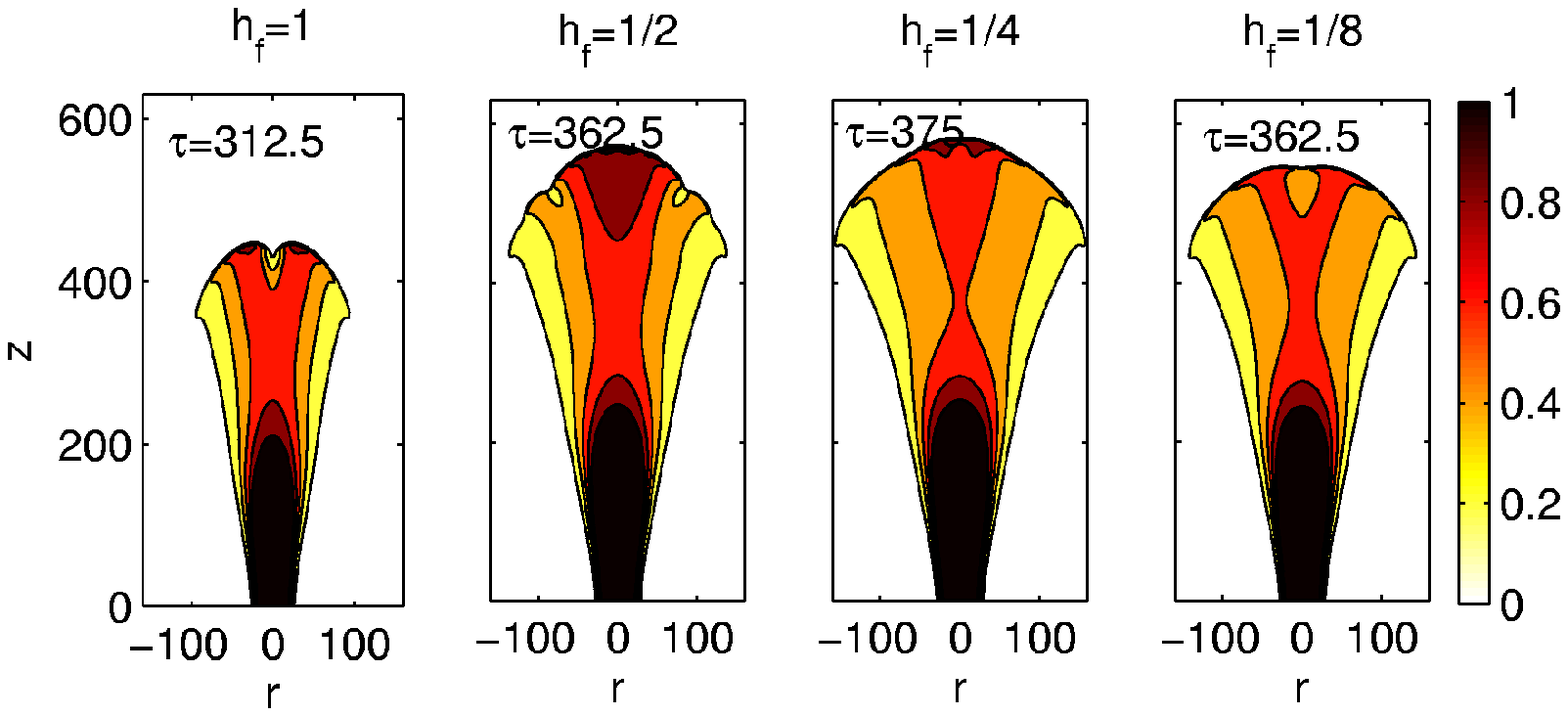}
\caption{Upper panel: Branching time in a background field $|\E_b|=0.5$ 
as function of the finest mesh size $h_f$ = 2, 1, 1/2, 1/4, 1/8.
Lower panels: the corresponding electron density distribution at $\tau$=275 
(middle row), and just after the respective branching time  (lower row), 
computed on different finest grids $h_f=$ 1, 1/2, 1/4 and 1/8. 
The total computational domain is $z\le L_z= 2048$ and $r\le L_r = L_z/2$.}
\label{tsplits}
\label{sigma_tsplits}
\end{center}
\end{figure}

Adequate refinement criteria for the continuity and for the Poisson 
equation then lead to a grid distribution which is especially
designed to cope adequately and efficiently with the difficulties
inherent to both type of equations. More specifically, the refinement 
criterion for the grids for the Poisson equation is based on error estimate 
of the solution. The refinement criterion for grids for the continuity 
equations uses a curvature monitor of the solution. Moreover, it takes
explicitly into account the {\em leading edge},
where the densities are low but 
the electric field is greatly enhanced~\cite{ebe1997,ebe2000}.

The refinement criterion is computed at each  time step, in such a way that
the series of nested, consecutively refined grids move with the solution.
Special care has been taken for the discretizations as well as the mapping
of the solution from one grid to the other to be charge conserving. 
\\

\begin{figure}
\begin{center}
\includegraphics[width=8cm]{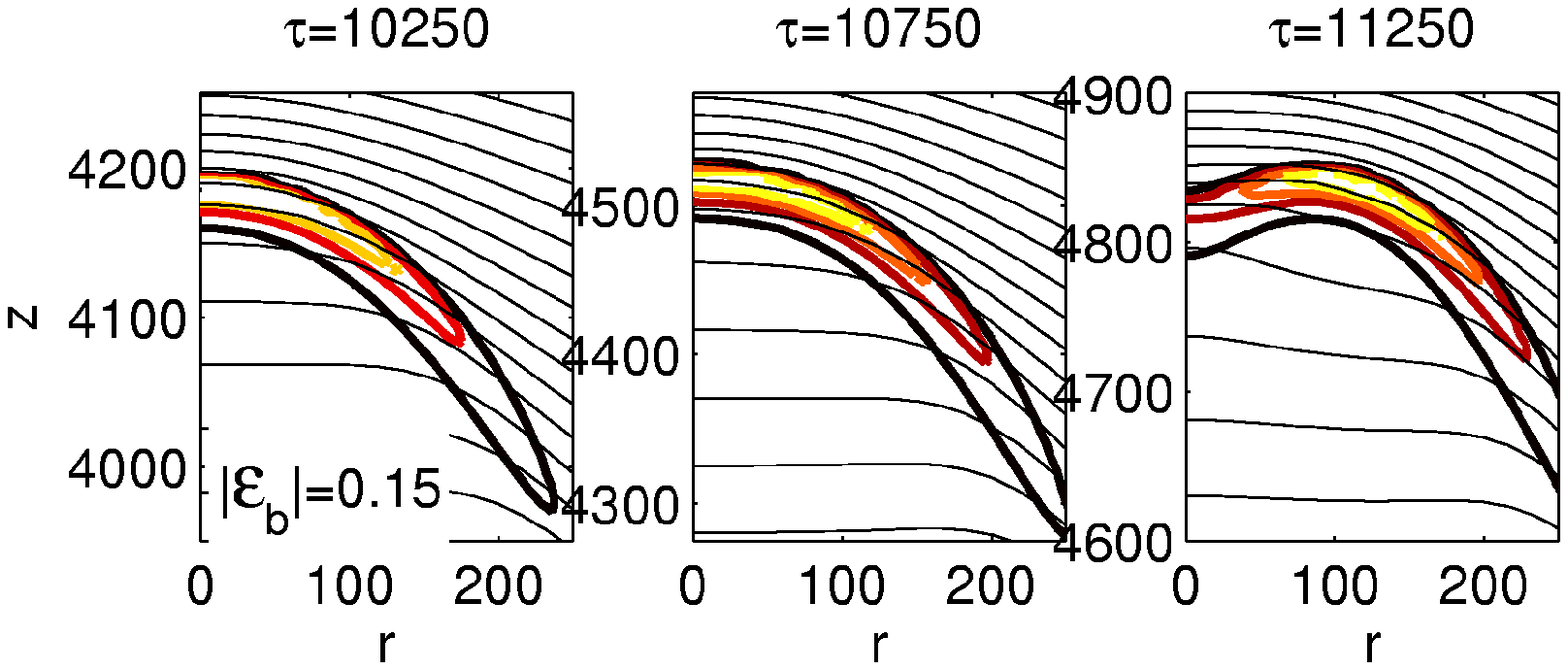}
\includegraphics[width=8cm]{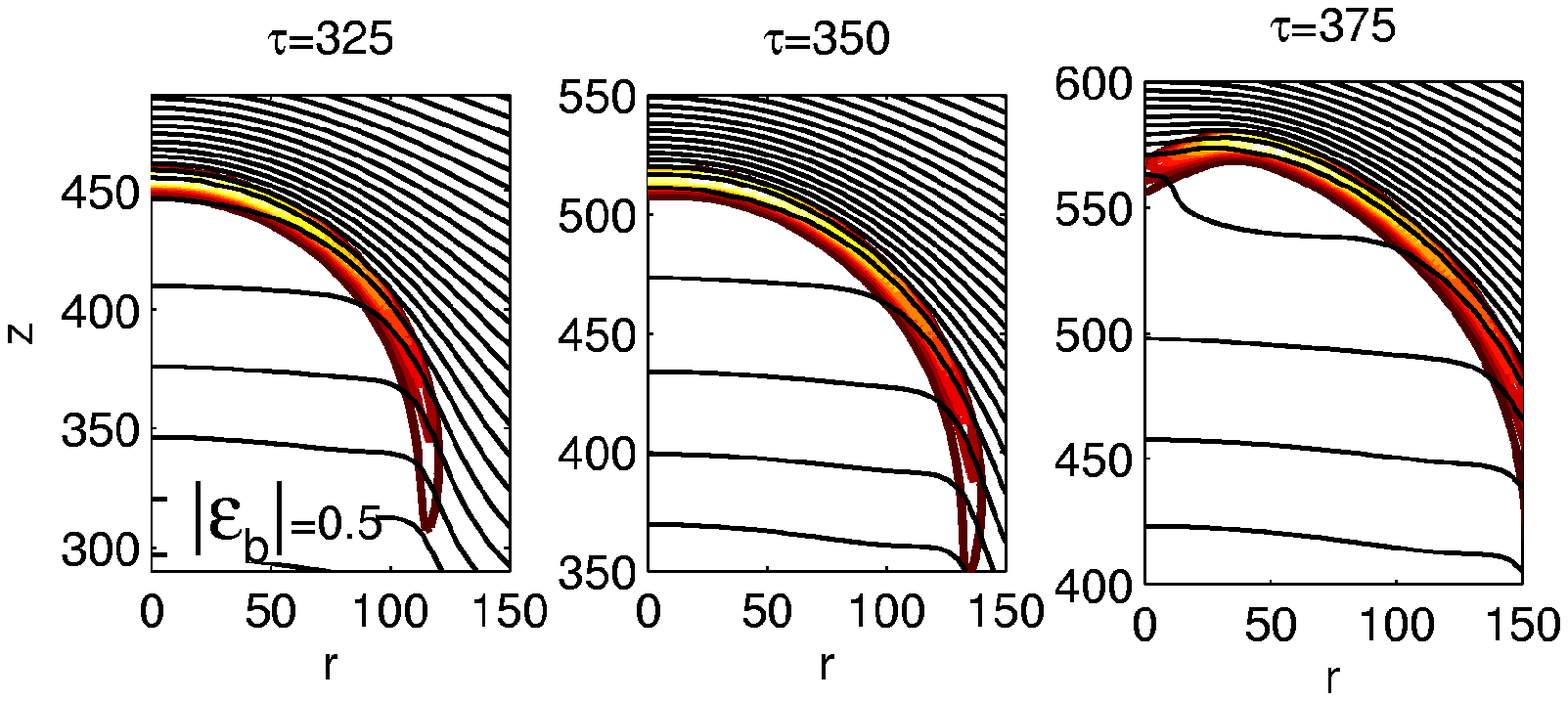}
\caption{Zoom into the streamer head during branching. Upper plots:
$|\E_b|=$0.15 as in Fig.~\ref{br015}, $h_f$ = 2. Lower plots: 
$|\E_b|=$0.5 as in Fig.~\ref{tsplits}, $h_f$ = 1/8.
Contour lines (thick) of net charge density and equipotential lines 
(thin) are shown as a function of positive radius $r$ and appropriate 
$z$. 
The spacing of the charge contour levels is 0.004 for the low field case, 
and 0.16 in the high field case. The spacing of equipotential lines
is 5 in both cases.} 
\label{chargephibr}
\end{center}
\end{figure}

{\bf Results.}
The adaptive grid refinement procedure enables us to resolve the streamer
with very high accuracy, and thus to investigate the dependence
of the branching process on the numerical grid.
The results are obtained
on increasingly finer grid sizes $h_f$, always taking the same coarsest
mesh width $h_c$ for both the continuity and the Poisson equations. If
the branching were of numerical nature, we would
expect that branching times on increasingly finer grids would not 
converge.
\\

{\it We first consider negative streamers evolving in a low background 
field of $\E_b=0.15$ corresponding to 30 kV/cm for N$_2$ at atmospheric 
pressure.} We use an electrically neutral, dense and relatively wide 
Gaussian ionization seed at the cathode, with a maximum of 1/4.8, and a 
characteristic radius 
of 10. This corresponds to a maximal electron and ion number density 
of 10$^{14}$ cm$^{-3}$ and an 1/e radius of 230 $\mu$m.
The gap length and width are set to $L_z=2L_r=2^{15}=32768$, which 
corresponds to an inter-electrode distance of approximately 7.5 cm.

The coarsest mesh width is set to $h_c=64$, and the finest
one to $h_f=$ 8, 4 and 2.   When a finest mesh of 8 is used, the electron
density in the streamer is lower, as can be seen in the
upper row in Fig.~\ref{br015}. This is due to the numerical
diffusion introduced on such a coarse grid by the flux limiter
that switches to the diffusive first order scheme in regions with large
gradients. This numerical diffusion smears the electrons out over the
streamer head, which in turn results in lower field enhancement
and lower ionization rates. The results on finer meshes of 4 and 2
on the other hand do agree with each other.

The branching in time is the same in all cases.
Fig.~\ref{br015} shows that the influence of the numerical grid on
the branching state rapidly decreases, and we thus 
can carry out not only qualitative but also quantitative numerical 
experiments of the streamer evolution up to branching. These results
show that branching is possible at lower electric fields than those
of~\cite{arr2002,roc2002}.
Branching was not observed in earlier simulations at lower 
fields~\cite{dha1987,vit1994} because the discharge gap was too short.
\\

{\it We now consider a negative streamer in a dimensionless background
field of $\E_b=0.5$ corresponding to 100 kV/cm in N$_2$ at atmospheric
pressure} in a gap of $L_z=2048$, or 4.6 mm. These external parameters 
are as in~\cite{arr2002,roc2002}. 
The initial seed is also taken as in~\cite{roc2002}, i.e., a Gaussian 
with amplitude 10$^{-4}$ and characteristic radius 10,
which corresponds to a maximal electron and ion number density of
$5\cdot10^{10}$ cm$^{-3}$ and an 1/e-radius of 23 $\mu$m for N$_2$ under
normal conditions. 

However, while \cite{arr2002} used a 
uniform grid of $h=2$ and \cite{roc2002} one of $h=1$, we now perform 
computations on a finest grid as small as $h_f=1/8$, i.e., 
more than a decade finer. More precisely,
the coarsest mesh width is set to $h_c=2$, and the finest
one to $h_f=2,1,\ldots,1/8$. Furthermore a better numerical scheme is used:
flux limiting~\cite{mon2005-2} rather than 3rd order 
upwind~\cite{arr2002,roc2002}.

Before branching, at $\tau$=275, Fig.~\ref{tsplits} shows that 
there is a quantitative difference between the results on a mesh with 
$h_f$=1 and the other three. As in the low field case, numerical 
diffusion spreads the space charge layer, which makes the field 
enhancement at the streamer tip and the field screening in the streamer 
body less efficient. Consequently, the ionization rate, 
and therefore the electron density, are higher in the 
streamer body. In the low field case we do not observe this because 
the background field is negligibly low, hence a less efficient screening 
will not affect significantly the ionization rate in the streamer body. 
It is clear that on meshes finer than 1/2, the results are the same 
during the stable streamer propagation. It is only after the branching 
that different states are observed on those very fine grids.
However, the time of branching converges within this range of mesh widths
$h_f$ as shown in the upper plot in Fig.~\ref{tsplits}. 
\\

{\bf Discussion, conclusion and outlook.}
We emphasize that the branching times converge on decreasing 
numerical grids in both cases. Therefore we here present the first 
conclusive and quantitative numerical predictions on streamer branching.
However, in contrast to the low field case, 
the lower plots in Fig.~\ref{sigma_tsplits} show that in 
the high field case different branched modes are reached after
approximately the same evolution time: in two cases,
the maximal electron density and field is on the axis 
of symmetry, and in two other cases, it is off axis. Apparently,
there are different branched states reachable at bifurcation
and tiny differences determine which one will be reached.
Such extreme sensitivity is well-known from deterministic chaos;
it is generic for nonlinear dynamics near bifurcation points.
On the other hand, the unstable state is reached in a
deterministic manner, and therefore the branching times converge.

%
But why is there once a unique branched state and once several?
The answer can be found in Fig.~\ref{chargephibr} showing the
two relevant spatial scales, namely
the thickness of the space charge layer and the radius of
the channel. In the high field case, the ratio of layer thickness
over radius is much smaller than in the low field case. Moreover,
the field screening and enhancement is much stronger 
and the equipotential lines follow the space charge layer much better.
Therefore the high field streamer is much closer to interfacial
models as discussed in~\cite{ebe1997,arr2002,meu2004,meu2005,ebe2006}
and can access more branching modes. This critical state
in future work will be characterized by the electric charge content
and electric field and potential at the streamer tip which would then 
allow us to relate branching to the external electric circuit. 
For sketches of such ideas as well as for a discussion 
of photo-ionization effects and of continuum versus particle models,
we refer to \cite{ebe2006}. 

{\bf Acknowledgment:} C.M.\ thanks the Netherlands 
Organization for Scientific Research (NWO) for a Ph.D.\ grant 
within the program on Computational Science.


\raggedbottom
\end{document}